# Kinetics of (2×4) → (3×1(6)) structural changes on GaAs(001) surfaces during the UHV annealing

A.V. Vasev[*], M.A. Putyato, V.V. Preobrazhenskii

*Institute of Semiconductor Physics, Siberian Branch of the Russian Academy of Sciences, Laboratory of Physical Bases of Semiconductor Heterostructures Epitaxy, Acad. Lavrentiev Avenue, 13, 630090, Novosibirsk, Russia*

ABSTRACT

The peculiarities of superstructural transition (2×4) → (3×1(6)) on the GaAs(001) surface were studied by the RHEED method in the conditions initiated by a sharp change of the arsenic flux. The specular beam intensities RHEED picture dependences on time were obtained during the transition. The measurement results were analyzed within the *JMAK* (*Johnson – Melh – Avrami – Kolmogorov*) kinetic model. It was established that the process of structural rearrangement proceeds in two stages and it is realized through the state of intermediate disordering, domains with different reconstructions being coexistent on the surface. The activation energies and phase transition velocities were determined for each of the stages. The procedure for precise determination of GaAs(001) surface temperature using the features of the $\alpha(2\times 4) \to DO$ transition process kinetic was proposed.

The results of this work allow us to broaden our understanding of the reconstruction transitions mechanisms. This information has a key (fundamental and applied) nature for the technologies of epitaxial growth of multilayer heterostructures, where the interface planarity and the sharpness of composition profile are of particular importance.

## 1. Introduction

Studying the nature of surface phase transitions is of fundamental and applied characters. This subject is of great scientific interest due to its importance for a wide circle of technological tasks. Understanding the reconstruction transition mechanisms at the microscopic level is of great importance for semiconductor compounds surfaces, but it is difficult to achieve because of the complexity of their nature.

A big number of experimental and theoretical contributions are devoted to the study of the GaAs(001) surface superstructural transitions properties. It is connected with the role they play in MBE processes [1-6]. Superstructural states affect the structural perfection of the obtained heterointerfaces, determining the characteristics of adsorption and desorption processes, and the diffusion of growth components on surface terraces, also the efficiency of their incorporation into a growing layer.

Of special interest, in this respect, is the superstructural transition from arsenic-rich reconstruction (2×4) to arsenic-poor reconstruction (3×1(6)). It is explained by that the conditions for the existence of this transition coincide with the region of GaAs(001) MBE growth conditions in which it is possible to obtain the most structurally perfect films [7].

Studying the structural peculiarities on crystal surfaces is most often realized with the Reflected High-Energy Electron Diffraction (RHEED) method. It is conditioned by a high informativity of this method that allows carrying out real time *in-situ* investigations of epitaxial growth and vacuum annealing processes. The RHEED-data contain the information about the structure and morphology of the surface under study.

The peculiarities of the superstructural changes on the GaAs(001) surface during the transition $\alpha(2\times 4) \to (3\times 1(6))$, initiated by a sharp change of the arsenic flux, were analyzed with the RHEED method.

---
[*] Tel.: +7 (383) 333 1967; fax: +7 (383) 333 3502.
 *E-mail address*: vasev@isp.nsc.ru.

## 2. Experimental

The experimental part of this work was carried out on the modernized $A^{III}B^{V}$ MBE "Shtat" machine.

The MBE machine is equipped with a valve-type arsenic source with a cracking zone. At the cracking zone temperature $T^{crack} = 950°C$ the composition of molecules in the flux is $BEP(\Sigma As) = 92\%(As_2) + 8\%(As_4)$ [8, 9].

The gallium atom flux was formed by a two-zone effusion cell.

An ionization vacuum Bayard-Alpert ionization gauge was used to determine the molecular fluxes density. The sensor was moved to the substrate position in the measurements regime.

The geometry of the cryopanels located around the growth zone and their temperature (cooled with liquid nitrogen) allowed an immediate and effective control of the $As_2$ molecular flux in the substrate plane. In particular, at closing the source valve, a 10-fold flux density decrease in the substrate position proceeds during 0.4 sec. Thereat, the background pressure in the growth zone equaled $3 \times 10^{-10}$ Torr. The total of these parameters guaranteed the absence of parasite influence of residual arsenic fluxes on the kinetics of the processes under study.

The electron beam energy at RHEED measurements was 25 keV. The observations were carried out in the azimuth [110] at electron beam incidence angles ($\alpha$) in the range from 0.46 to 2.42 of angular degree.

The samples were attached onto a molybdenum carrier with the help of melted indium and that provided the heterogeneity of temperature distribution over the surface. The substrate temperature ($T_S$) was controlled on the indices of the thermocouple fixed in the carrier material and that provided a direct thermal contact of the thermocouple with the sample. The correctness of thermocouple indices at determining the substrate temperature was controlled by the position of the transitions between surface superstructures $c(4 \times 4) \to (2 \times 3)$ and $(2 \times 3) \to \gamma(2 \times 4)$ on the GaAs(001) surface [10]. The investigations were realized in the range of substrate temperatures $T_S$ from 550° to 614°C.

The *epi-ready* substrates of semi-insulating GaAs(001) with a misorientation of ~8 angular minutes towards [110] were used for the studies. The samples were MBE-grown ~1 μm thick buffer layers of homoepitaxial GaAs. Before each measurement, the sample surface was renewed by means of overgrowing with a ~50 nm thick layer at $T_S = 580°C$. The arsenic flux used in the overgrowing guaranteed the formation of a superstructural state with clear symmetry (3×1) on the surface [7].

After the end of preparatory procedures the chosen $T_S$ and $\alpha$ values were set. The arsenic flux density had been chosen such that reconstruction $\beta(2 \times 4)$ was formed at a set $T_S$ value on the starting surface. Upon reaching the stationary state by the surface, the arsenic flux was closed and the change of RHEED picture specular beam intensity was registered on the time at superstructural transition $\alpha(2 \times 4) \to (3 \times 1(6))$.

## 3. Experimental results and discussion

The data of RHEED picture specular beam intensity behavior during superstructural transition $\alpha(2 \times 4) \to (3 \times 1(6))$ on the GaAs(001) surface at $T_S = 610°C$ and starting pressure $BEP(As_2) = 1.7 \times 10^{-6}$ Torr are presented in Fig. 1. The family of $a \div r)$ curves illustrates the evolution of characteristic *SBI* time dependence at a change of electron beam incidence angle $\alpha$ in the range from 0.46 to 2.42 of the angular degree (with step 0.12°). The indicated peculiarities of time dependences characterize the superstructural transitions influence on the ability of surface to interact with an electron beam. Thereat, it is necessary to remember that each curve of the $a \div r)$ family in Fig. 1 (despite their external difference) describes one and the same sequence of structural transition processes. In particular, the sequence and type of the processes that proceed on the surface, their velocities, times of beginning and ending are the same for all $a \div r)$ curves. As a consequence, if we assume that dynamical (multi-act) scattering effects do not have their dominant influence on the *SBI* formation, then the angular dependence in Fig. 1, apart from kinetic parameters, will have the information only about the structure of the states between which these transitions are realized.

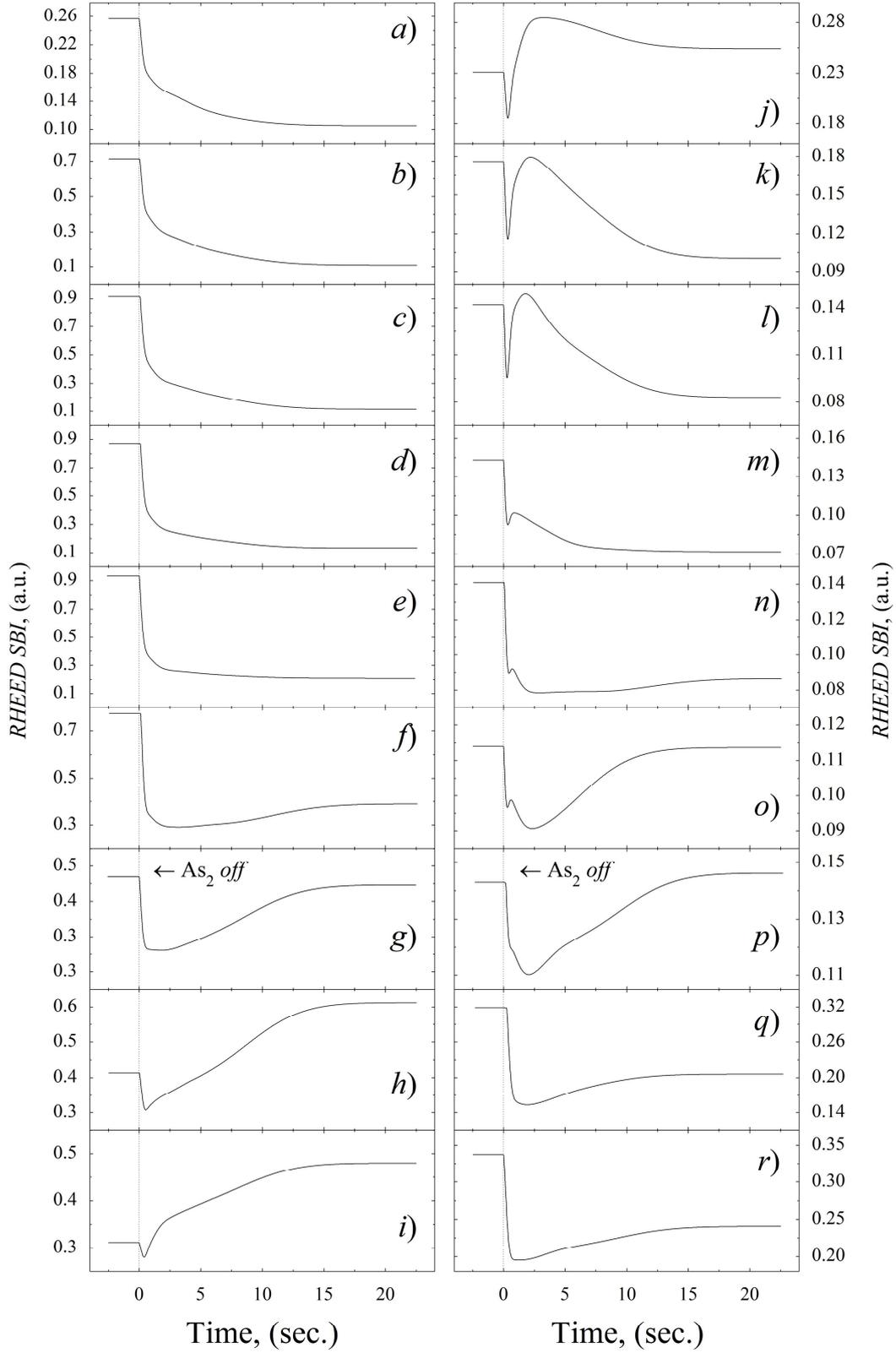

**Fig. 1.** Specular beam intensity evolution of the RHEED picture during superstructural transition $\alpha(2\times4) \to (3\times1(6))$ on the GaAs(001) surface. The transition is initiated by a sharp change of As$_2$ molecular flux value. Curves $a) \div r)$ correspond to the electron beam incidence angle values in the range from 0.46 to 2.42 of angular degree (with step 0.12°).

Analyze the characteristic features of the $a \div r)$ curves in Fig. 1 in the assumption that the following condition is fulfilled:

$$I^{SB} = \sum_j I_j \cdot \theta_j . \qquad (3.1)$$

Here $\theta_j$ – the degree of surface coverage by domains of superstructural states $j$, and $I_j$ – the RHEED $SBI$ that corresponds to a surface fully covered by superstructural state $j$. For a preliminary analysis, let us choose the curve $o)$ in Fig. 1 as the most topologically complicated one (has three clearly expressed extremums). It follows from expression (3.1) that such situation can be realized only on one condition: a sequence of superstructural conditions should consist of five states (Fig. 2).

One can state with a high degree of confidence, as regards states 1 and 5, that these are reconstructions $\alpha(2\times 4)$ and $(3\times 1)$, respectively. This conclusion is confirmed by both a visual observation of clear RHEED pictures and the registration of RHEED rocking curves typical of these reconstructions.

State 2 is realized on the surface as a result of the quick (~ 0.4 sec.) process. This process is accompanied by a blurring of RHEED picture fractional-order spots $\alpha(2\times 4)$ and a sharp increase of diffuse background. It is possible to state that a rapid closing of the arsenic source valve initiates an intensive desorption process of arsenic dimers from the sample surface. As a result, state 2 is rather an $\alpha(2\times 4)$ reconstruction with a big number of As-dimer vacancies. The surface has preserved its short-range order, but already lost its long-range order. Further, we will designate this state as "$DO$".

No other RHEED pictures, but the superposition of blurred spots $(0, 1/2)$ and $(0, 1/3)$ in azimuth [110] (symmetry of states 3 and 4 - unestablished) at the formation of states 3 and 4 on the sample surface, were observed. This evidences small fractional-order spot intensities, compared to a high diffuse background. As a consequence, we can judge about the structural properties of states 3 and 4 only on the RHEED rocking curves data.

The data of Fig. 1 can be presented in a way more convenient for visual perception: namely, as the RHEED rocking curves evolution in superstructural transition processes (Fig. 3). The rocking curves can be divided into four groups: $a)$ $1 \to 2$ $(0.1\ s \div 0.4\ s)$, $b)$ $2 \to 3$ $(0.4\ s \div 0.7\ s)$, $c)$ $3 \to 4$ $(0.7\ s \div 2\ s)$ and $d)$ $4 \to 5$ $(2\ s \div 20\ s)$. The chosen time intervals correspond to the structural transition intervals of the diagram of Fig. 2.

The validity of this choice follows from the analysis of the curves $e \div h)$ in Fig. 3, which illustrate the relative changes of RHEED rocking curves $\delta SBI \equiv I^{SB}_{current} - I^{SB}_{start}$.

The RHEED rocking curves for the superstructural states, realized in stationary conditions (at fixed temperature and arsenic fluxes) with the known symmetries: $a)$ $\alpha(2\times 4)$, $b)$ $(3\times 6)$ and $c)$ $(3\times 1)$, see Fig. 4, were also additionally obtained for the data analysis and interpretation of Fig. 3. The measurements were made at $T_S = 580°C$. The arsenic flux value was fitted from the consideration of forming the clearest RHEED picture with a set symmetry.

It is shown in Fig. 3 $a)$ and $e)$ that the process $1 \to 2$ is characterized by the clearly expressed changes localized in the range of $0.46 \div 1.38$ angular degrees for $\alpha$ values. These changes are a decrease of the structural peak amplitude with its maximum in a position $\sim 0.88°$, observed in the course of time. Comparing the data of Fig. 3 $a)$ and $e)$ to that of Fig. 4 $a)$, we conclude on the fairness of the already formulated hypothesis: the formation of strongly disordered $\alpha(2\times 4)$ (state "$DO$") during the desorption process from the surface of As-dimers.

The process $2 \to 3$ is characterized by the appearance of new structural components – a pair of peaks with their maxima in positions $\sim 1.5°$ and ~ 1.73°, respectively (Fig. 3 b) and f)). These peak amplitudes grow with time. The character of peak changes with its maximum in a position ~ 0.88° does not change: its amplitude continues to decrease. We are inclined to interpret these changes as a start of the short-range order transformation in the structure of $\alpha(2\times4)$ with a following formation of a new (but structurally related) superstructural state.

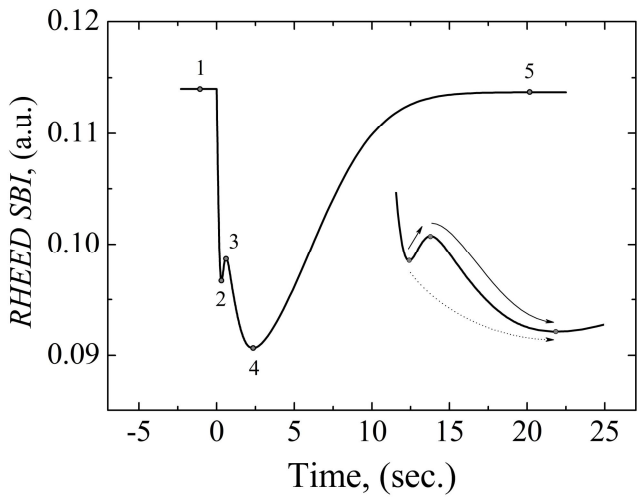

**Fig. 2.** Diagram of the superstructural states observed on the GaAs(001) surface during transition $\alpha(2\times 4) \to (3\times 1(6))$.

A more complicated picture of changes is typical of the process $3 \to 4$ (Fig. 3 c) and g)). In particular, the peak amplitude growth velocity with its maximum in a position ~ 1.5° becomes almost twice higher than that with its maximum in a position ~ 1.73°. Also, the peak with its maximum in a position ~ 0.88° is divided into two ones – with their maxima in positions ~ 0.69° and ~ 0.92°, respectively. Thereat, the decrease of the left peak amplitude velocity is ~ 1.3 times higher than that of the right one. Note that the peak formation with its maximum in a position ~ 0.69° indicates the presence of structural elements of state $(3 \times 6)$ (Fig. 4 $b$)) in the state 4 structure.

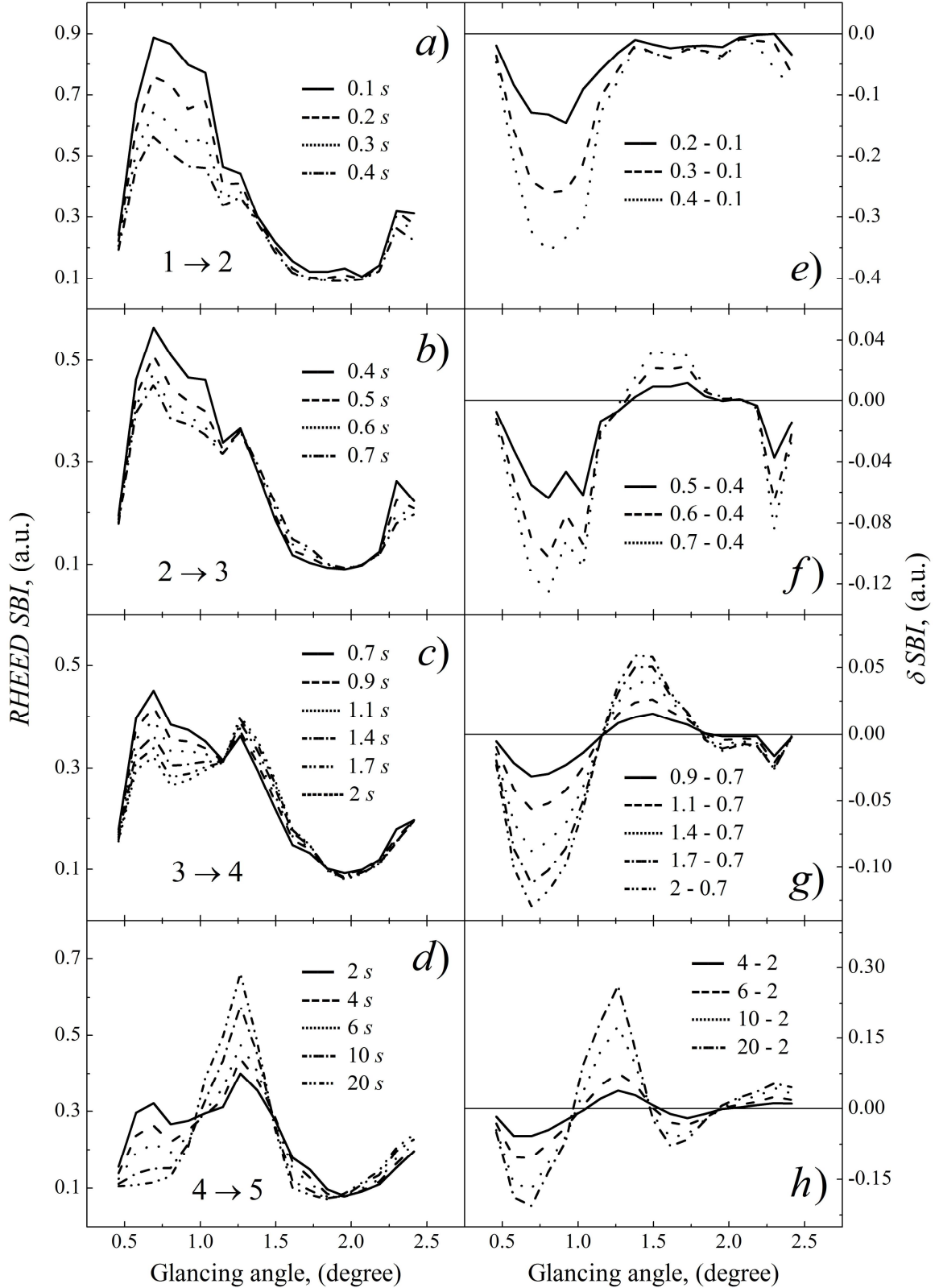

**Fig. 3.** Evolution of RHEED rocking curves during the superstructural transition $\alpha(2\times 4) \to (3\times 1(6))$ realized in the Langmuir desorption mode.

Finally, process $4 \rightarrow 5$ is the formation of reconstruction $(3 \times 1)$ (Fig. 3 $d$) and $h$)). It is seen in the figure that there forms a new structural peak with its maximum in a position ~ 1.27°. This indicates the appearance of structural components reconstruction $(3 \times 1)$ on the surface (Fig. 4 $c$)). Besides, a decrease (to zero) of the peak amplitude with its maximum in a position ~ 0.69° is observed, the peak amplitude with its maximum in a position ~ 0.92° being practically unchanged. The analogous picture is observed for the peaks with their maxima in positions ~ 1.5° and ~ 1.73°. The left peak amplitude is decreased to zero, and it remains practically unchanged with the right one.

The absence of experimental data on the symmetry of states 3 and 4 considerably complicates interpreting the data of Fig. 3 $f$) and $g$)). Nevertheless, we can be based on a number of indirect experimental and literary data.

First of all, we would like to point out the contribution by D. Martrou *et al.* [11], in which the authors, using the high-resolution STM method, showed that the reconstructed transition $(2 \times 4) \rightarrow (3 \times 1)$ on the GaAs(001) surface is a sequence of $(2 \times 4) \rightarrow (6 \times 6) \rightarrow (1 \times 1)_{45\%} + (3 \times 2)_{39\%} + (6 \times 6)_{16\%}$. Here $(1 \times 1)$ means that the surface is in the structurally unmodified state between domains $(3 \times 2)$ and $(6 \times 6)$. That is the atoms of top surface layer do not form dimer bonds and preserve their bulk lattice geometry.

The other contribution interesting for us can be the article by I. Chizhov *et al.* [12], where the authors, using the high-resolution STM method, showed that reconstruction $(3 \times 6)$ is a complex of $(2 \times 6) + (1 \times 1) + (3 \times 3)$. The authors also set forth a hypothesis that, due to the external similarity, $(3 \times 3)$ may be one of the variants of the state $(3 \times 2)/(3 \times n)$ earlier found by L. Li *et al.* [13].

As reconstruction $(3 \times 6)$ is formed between states $\alpha(2 \times 4)$ and $(3 \times 1)$ in stationary conditions (at a step reduction of arsenic molecular flux), one can assert that the main structural components of stationary reconstruction $(3 \times 6)$ may be present also in superstructural states 3 and 4. Such assertion is based on the idea of the "*inheritance*" of the main structural components at superstructural transitions, reported by I. Chizhov *et al.* for the transition model $(2 \times 4) \rightarrow (4 \times 2)$ on the GaAs(001) surface [12]. The RHEED rocking curves evolution in Fig. 3 visually confirms the validity of using this model.

Generalizing the above-formulated, it is possible to set forth a supposition that we observe $\alpha(2 \times 4) \rightarrow (6 \times 6)$ as transition $2 \rightarrow 3$, and we register $(6 \times 6) \rightarrow (1 \times 1) + (n \times 6)$ as transition $3 \rightarrow 4$. It will be suitable to note here that designation "$(1 \times 1) + (n \times 6)$" does not characterize the symmetry of the observed RHEED picture, but it carries the information about the type of the structural components present on the surface. Further, we will designate state 4 as "$(n\times 6)$". The symmetry of the RHEED picture we observe at the transition is indicated as "$(3\times 1(6))$".

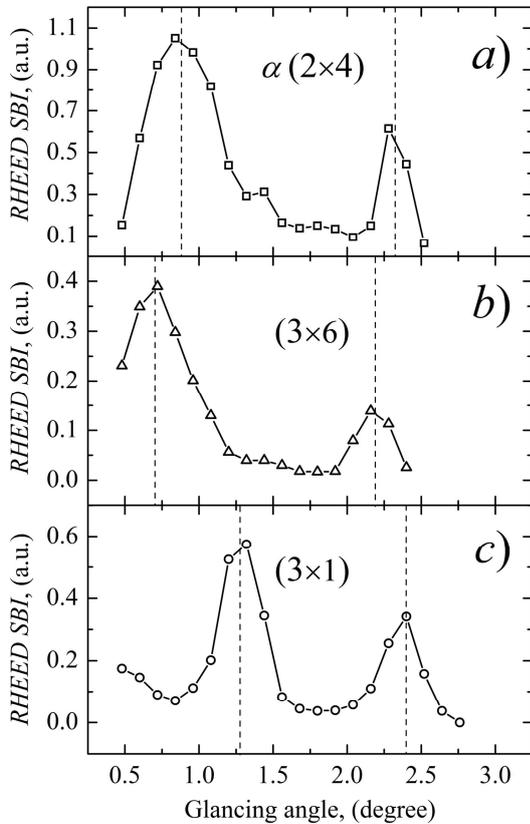

**Fig. 4.** Rocking curves RHEED of the superstructural states *a*) $\alpha(2 \times 4)$, *b*) $(3 \times 6)$ and *c*) $(3 \times 1)$ for the GaAs(001) surface obtained in stationary conditions. □, △ and ○ – exp. data.

We would like to stress out the fact that the term "*structural peak*" we use at analyzing the data of Fig. 3 describes a concrete physical process: the result of electron diffraction on an element of the surface structure, which periodicity becomes "*visible*" for electrons just in the range of $\alpha$ angles, where this structural peak is observed. As a consequence, the structural peak is considered by us as a whole object (in the angles range of its existence) that contains information about surface structural properties. A considerable conclusion from this supposition is the statement about the identity of the structural information obtained at measurements for different angle values $\alpha$ from the structural peak existence range.

As a surface RHEED rocking curve can be considered as a result of all structural peaks superposition in all the investigated range of angles $\alpha$, one can point out some possible special sites of this range.

First of all, these are the sites where the existence ranges of peak angles cross each other. Consider the situation when the mentioned structural peaks characterize the properties of differing superstructural states (in the simplest case – the initial and finite states of the superstructural transition). Then, the measurements made in such conditions (at the cross of existence ranges) allow obtaining more complete information about the key members of the transition.

Those ranges of angles $\alpha$, where the RHEED picture *SBI* undergoes maximal changes during a superstructural transition (to improve the *signal/noise* ratio), are also preferable.

Finally, the considered angles range should satisfy the condition when the specular baem position in the RHEED picture is maximally distant from artifacts having a dynamical (multi-act) scattering nature.

The RHEED rocking curve region that satisfies all the three above-described conditions is called "*sensitivity window*". It follows from the analysis of Fig. 3 that the range of $1.0 \div 1.5$ angular degrees is such a window for transition $\alpha(2\times 4) \rightarrow (3\times 1(6))$.

The angle $\alpha$, chosen to realize the measurements, should satisfy the diffraction geometry that guarantees the minimal RHEED signal sensitivity to the diffuse components, caused by a scattering on $\leq 1$ ML high objects. The fulfilment of the latter condition is easily controlled on the parameters of RHEED picture spots (they should be maximally narrow, short and clear). As a result, the experimental conditions at $\alpha = 1.36°$ are the most optimal.

When carrying out a quantitative analysis of experimental RHEED data, it is necessary to take into account the fact that the RHEED-registered features for the existence of state 3 have a value (on the amplitude) ~ 0.005 a.u. (Fig. 3 *m)*, *n)* and *o)*). This value is close to the sensitivity limit of the equipment. It is also important that the indicated features are observed only in the narrow ($1.84 \div 2.07$ angular degrees) range of $\alpha$ values. As a consequence, to simplify the model description, it is natural to neglect the fact of the existence of state 3. In other words, one transition $2 \rightarrow 4$ (dashed line in the diagram of Fig. 2) was considered instead of the complex of transitions $2 \rightarrow 3 \rightarrow 4$.

The fairness of this approach is illustrated in Fig. 5, where the results of the model description for the experimental data of curves *d)*, *i)*, *k)*, *o)* and *q)* (Fig. 1) are presented. According to the model, the RHEED picture *SBI* is described by the expression (3.1), where the time dependences of coverage degrees $\theta_j$ are presented in Fig. 6. The RHEED rocking curves for states $\alpha(2\times 4)$, *DO*, $(n\times 6)$ and $(3\times 1)$ (Fig. 7) are used as weighting coefficients $I_j$.

It is seen in Fig. 5 that the model built on the base of expression (3.1) accurately describes both *SBI* evolution curves proper and their time derivatives $\partial I^{SB}/\partial t$ in all the range of $\alpha$ angles we use. The artefacts connected with state 3 (presented in more details in the inserts to curves *d)* and *i)* in Fig. 5) are not the features of the incorrectness of expression (3.1), but they reflect a conscious simplification at excluding state 3 from the consideration. It is necessary to emphasize that expression (3.1) is a generalized form of recording the RHEED rocking curves additivity condition, and Fig. 5 is an illustration of fulfilling this additivity in all the range of $\alpha$ angles we use. Besides, it is an indirect, but a most considerable argument in favor of the supposition about the insignificance concerning the influence of dynamical (multi-act) scattering effect. This influence is confined to the formation of the absolute weighting coefficient values $I_j$ in expression (3.1) and, as a consequence, it does not affect the procedure of determining the kinetic and thermodynamic parameters of the analyzed processes.

It is worth pointing out the fact that our investigations of the kinetic and thermodynamic parameters of structural transition $\alpha(2\times4) \to (3\times1(6))$ were realized in the substrate temperature range of 570 ÷ 614°C. As a consequence, the presence of superstructural state $(3\times1)$ on the surface at low temperatures was negligibly small. Thus, only the transitions between three states $\alpha(2\times4)$, $DO$ and $(n\times6)$ were taken into account.

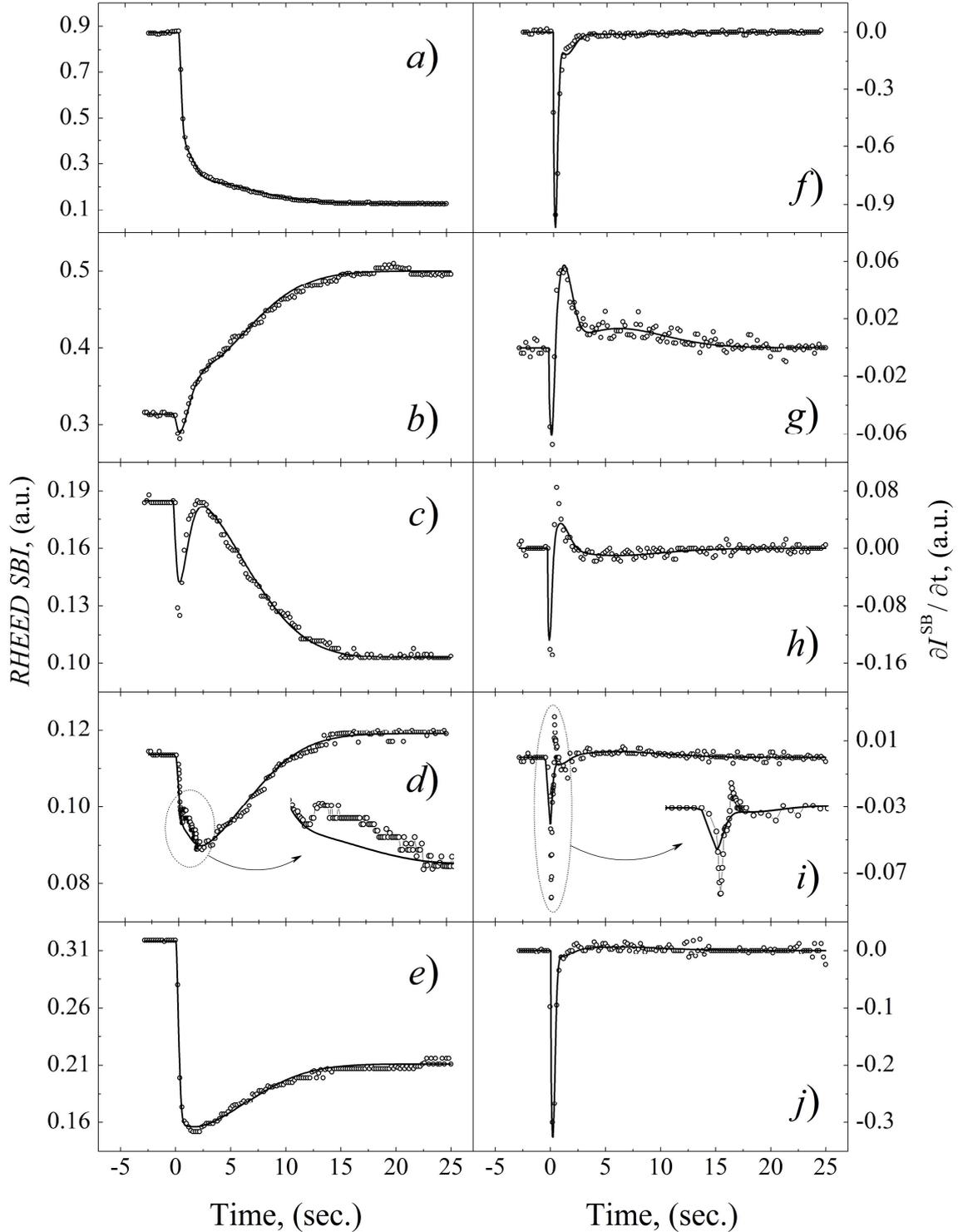

**Fig. 5.** Model description of RHEED picture *SBI* evolution during superstructural transition $\alpha(2\times4) \rightarrow (3\times1(6))$ for the $\alpha$ angles equal to $0.81°$ – *a*) and *f*), $1.38°$ – *b*) and *g*), $1.61°$ – *c*) and *h*), $2.07°$ – *d*) and *i*), $2.3°$ – *e*) and *j*), respectively.

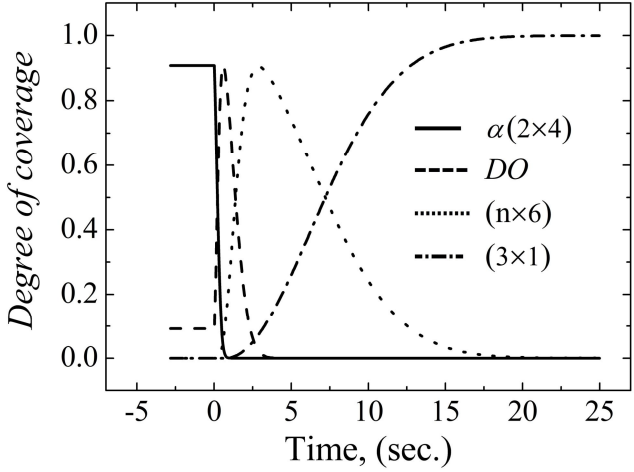

**Fig. 6.** Change of the degree of GaAs(001) surface coverage by the domains of superstructural states $\alpha(2\times4)$, $DO$, $(6\times6)$, $(n\times6)$ and $(3\times1)$ during transition $\alpha(2\times4) \to (3\times1(6))$.

The registration of RHEED picture fractional-order spots intensity evolution during a superstructural transition is one of the ways to check the model fairness based on expression (1.1). In particular, the fractional-order spots (0, 1/2) and (0, 1/3) in azimuth [110] can be used for these purposes. These spots are spatially separated in the RHEED picture and characterize the noncoinciding types of superstructural states symmetries.

According to the experimental data, the spot intensity (0, 1/3) during transition $\alpha(2\times4) \to DO$ did not exceed the diffuse background level. As a consequence, the influence of the structural components of state $DO$ on the spot (0, 1/3) can be thought of as negligibly small. That is the fractional-order spot (0, 1/3) is an individual characteristic of superstructural state $(3\times1(6))$.

Obtaining the quantitative information about the parameters of superstructural transition from the data of RHEED picture fractional-order spots evolution is possible only within some model views about the system under study.

Lower is our assumption for building such model description:

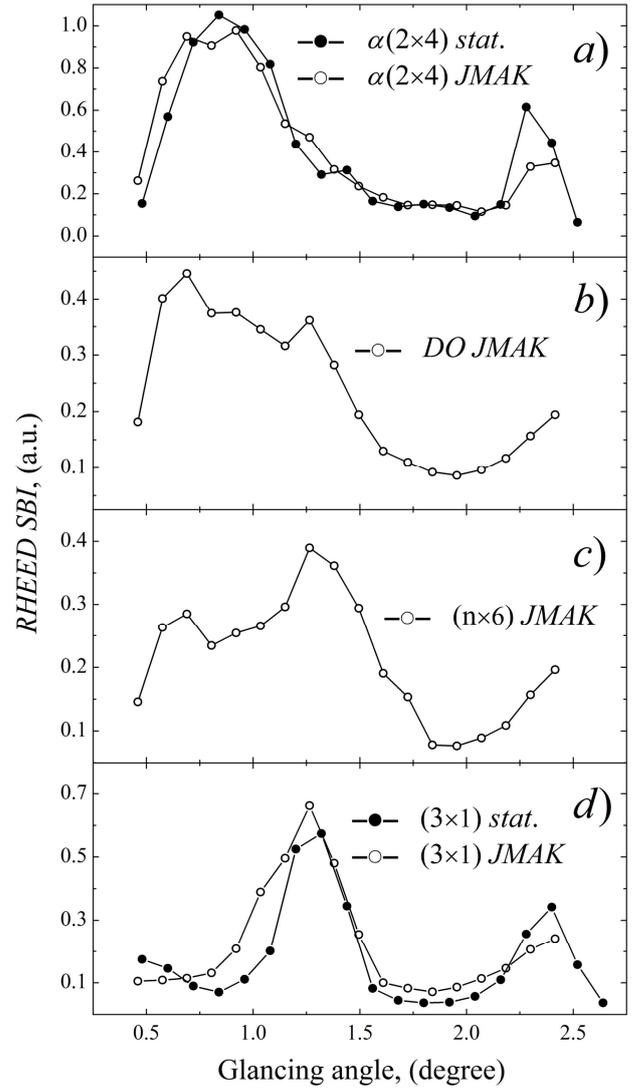

**Fig. 7.** RHEED rocking curves of basic superstructural states $\alpha(2\times4)$, $DO$, $(6\times6)$, $(n\times6)$ and $(3\times1)$ in the description model of transition $\alpha(2\times4) \to (3\times1(6))$.

*At a superstructural transition between two states, minimum one of them should satisfy the definition of Gibbs phase, i.e. it should be a homogeneous part of a heterogeneous system with a clearly expressed border.*

In other words, the island (domain) mechanism of superstructural transition should be realized in a studied system. A domain should be the state of a deep minimum of free Gibbs energy for a surface system. That will provide fulfilling the condition of composition homogeneity and structural properties (short-range and long-range order) of the surface inside domains during the whole superstructural transition. Besides, it will require localizing all the processes connected with a change of surface composition and structure in the front zone of this transition (at domains borders).

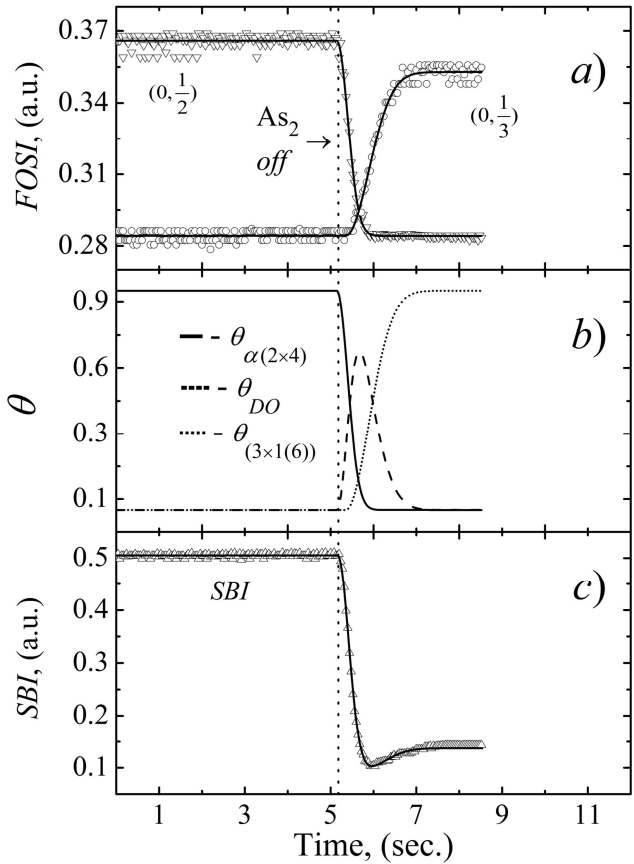

**Fig. 8.** Change of *a)* fractional-order spots and *c)* specular beam intensities at superstructural transition $\alpha(2\times4) \to (3\times1(6))$; *b)* restored time dependences of the degrees of GaAs(001) surface coverage by the domains of superstructural states $\alpha(2\times4)$, *DO*, and $(3\times1(6))$.

○, ▽ and △ – exp. data; — – fitting.

The Si(111) surface can serve as example of such system during superstructural transition $(7\times7) \to (1\times1)$. It was studied in detail and described in literature [14-17].

For this system, K. Shimada *et al.* [18] showed the principal possibility of using the data on RHEED picture spots intensity evolution for a quantitative determination of the degree of surface coverage by reconstruction domains. The investigations were the registration of spots (0, 0), (0, 1/7) и (3/7, 4/7) intensity changes in real time and a parallel measurement of the degree of Si(111) surface coverage by domains with reconstruction (7×7) with the high-temperature STM method. It was established that these parameters are linearly dependent. Moreover, the authors showed that the evolutions of normalized spots (0, 0), (0, 1/7) and (3/7, 4/7) intensities coincide during the whole transition $(7\times7) \to (1\times1)$. It allowed the authors to state that the indicated fractional-order spots of the RHEED picture carry the same information about the degree of surface coverage by reconstruction domains (7×7).

The results obtained by S. Hasegawa *et al.* in [19] confirm and enhance this assertion. The authors showed the coincidence of normalized intensities evolution for 8 more RHEED picture fractional-order spots of reconstruction (7×7): (2/7, 2/7), (3/7, 3/7), (4/7, 4/7), (5/7, 5/7), (6/7, 6/7), (4/7, 5/7), (5/7, 6/7) and (6/7, 1).

It can be concluded from the analysis of the results above that such picture will be typical of any spot of the RHEED pattern.

As other "standard" systems, one can mention the superstructural transitions $(\sqrt{3}\times\sqrt{3}) \to (\sqrt{31}\times\sqrt{31}) \to (4\times1)$ initiated by the deposition of submonolayer number of In atoms on the Si(111) surface at $T_S$ = 450°C [19]. The mentioned transitions are characterized by a linear (in time) change in the intensity of the spots of RHEED pattern during the exposition to a constant flux of In atoms. As the amount of In grows on the surface at the direct ratio to the exposition time, it is also indicative of the linear dependence of the spots of RHEED pattern intensity on the degree of surface coverage by the reconstruction domains in this system. Besides, the above-mentioned transitions are characterized by the ratio

$$\theta_A + \theta_B = 1, \qquad (3.2)$$

where $\theta_A$ and $\theta_B$ – degrees of surface coverage by domains with superstructural states *A* and *B*, respectively. Expression (3.2) is just for the transition $(\sqrt{3}\times\sqrt{3}) \to (2\times2)$ initiated by the deposition of a submonolayer number of In atoms on the Si(111) surface $\sqrt{3}\times\sqrt{3}$ - In at $T_S$ = RT [19].

The intensity changes of fractional-order spots (0, 1/2) and (0, 1/3) in Fig. 8 *a)* reflect the change in the degree of surface coverage by reconstructions $\alpha(2\times4)$ and $(3\times1(6))$, respectively. The data of coverage degree $\theta_{\alpha(2\times4)}$ and $\theta_{(3\times1(6))}$ are presented in Fig. 8 *b)*. The character of these changes during the transition $(\theta_{\alpha(2\times4)} + \theta_{(3\times1(6))} \neq 1)$ indicates the presence of the third structural state on the surface. The degree of the surface coverage by this state $\theta$ is determined by expression

$$\theta = 1 - \theta_{\alpha(2\times4)} - \theta_{(3\times1(6))} \qquad (3.3)$$

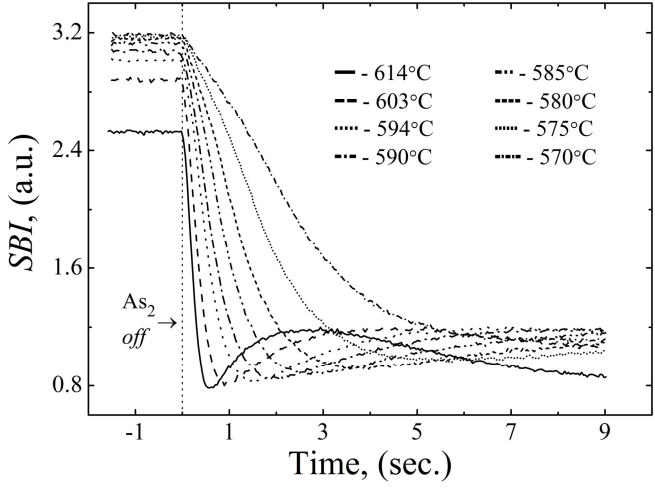

**Fig. 9.** *SBI* evolution during the superstructural transition at a change of substrate temperature in the range of 570 ÷ 614°C.

Value $\theta$ describes the degree of surface coverage by superstructural state *DO*. This conclusion completely agrees to the results of the data analysis of Fig. 3.

The *SBI* evolution, measured simultaneously with fractional-order spots intensity evolution, was presented as a linear combination of functions $\theta_{\alpha(2\times4)}$ and $\theta_{(3\times1(6))}$ with the corresponding weighting coefficients, (see Fig. 8 *c*)):

$$I^{SB} = \theta_{\alpha(2\times4)} \cdot I_{\alpha(2\times4)} + \theta_{(3\times1(6))} \cdot I_{(3\times1(6))} + \left(1 - \theta_{\alpha(2\times4)} - \theta_{(3\times1(6))}\right) \cdot I_{DO}. \quad (3.4)$$

In this case, the weighting coefficients are the specular beam intensities $I_{\alpha(2\times4)}$, $I_{(3\times1(6))}$ and $I_{DO}$ for the GaAs(001) surface in superstructural (2×4), (3×1(6)) and *DO*, respectively. It is seen in Fig. 8 *c*) that the calculated *SBI* curve completely coincides with the experimental data. Hence, *SBI* carries the same information about the degree of surface coverage by domains with different superstructural states as fractional-order spots intensities do.

Besides, when realizing the superstructural transitions monitoring, fractional-order spots intensity often falls down to background values. From this viewpoint, the specular beam has the best characteristics in its *signal/noise* parameter. As a consequence, studying the superstructural transitions kinetics is more convenient to realize controlling the *SBI* parameter.

The *SBI* evolution during superstructural transition $\alpha(2\times4) \to (3\times1(6))$, at the change of substrate temperature in the range of 570 ÷ 614°C is presented in Fig. 9.

The analysis of these data was realized within the *JMAK* (*Johnson-Melh-Avrami-Kolmogorov*) kinetic model [20-22]. The appropriateness of applying this model is confirmed by the character of the obtained results that allow considering the superstructural surface rearrangement as a process of new phase islands nucleation and their following growth and coalescence. According to this model, the change of a new phase fraction in the course of time *t* – in case when the phase transition is realized by the nucleation and growth of nuclei – is described by the expression:

$$\theta_{new} = 1 - \exp\left(-\gamma^d (t - t_0)^m\right) \quad (3.5)$$

where $\theta_{new}$ – the degree of filling the system bulk by a new phase, $\gamma$ – the parameter characterizing the phase transition velocity, $t_0$ – the moment of its onset, $d$ – the dimensionality of a system in which a transition is realized. If the nucleation velocity of nuclei in a bulk unit is constant, then $m = d+1$; if the number of nuclei is unchangeable during a phase transition, then $m = d$.

In our case, we deal with the initially 2D system (all objects are located on the surface and are characterized as $d \leq 2$). The domains of superstructural states $\alpha(2\times4)$ and *DO* act as the phases at transition $\alpha(2\times4) \to DO$ and, at transition $DO \to (3\times1(6))$ these are the domains of superstructural *DO* and $(3\times1(6))$.

At building the *JMAK*-description of transition $\alpha(2\times4) \to (3\times1(6))$ it was supposed that

$$\theta_{\alpha(2\times4)} = \begin{cases} \exp\left(-\gamma_{DO}^d \cdot (t_{off} - t_{DO})^m\right), & t < t_{DO} \\ \exp\left(-\gamma_{DO}^d \cdot (t - t_{DO})^m\right), & t_{DO} \leq t \end{cases} \quad (3.6).$$

Here $t_{off}$ is the moment of closing the arsenic source. Parameter $t_{DO}$ is of the *negative* values describing the presence of some starting amount of superstructural state *DO* on the starting surface (being in the stationary state under the arsenic flux). Such situation is typical for high substrate temperature values.

We have the following expression for the degree of surface coverage by superstructural state $(3\times1(6))$:

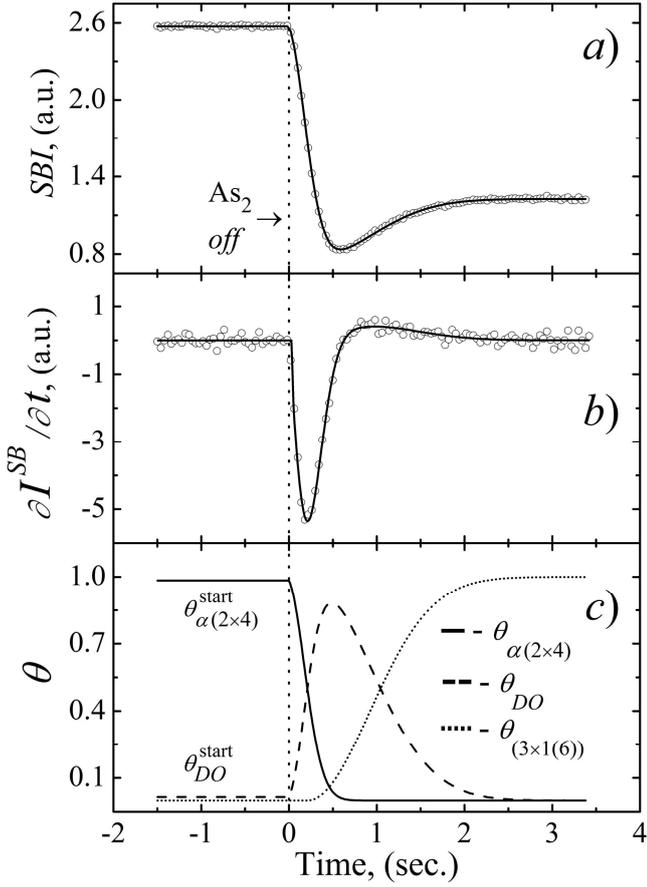

**Fig. 10.** Analysis of *SBI* evolution within the *JMAK* kinetic model.
○ – exp. data; ▬ – fitting.

$$\theta_{(3\times1(6))} = \begin{cases} 0, & t < t_{(3\times1(6))} \\ 1-\exp\left(-\gamma_{(3\times1(6))}^{d} \cdot \left(t-t_{(3\times1(6))}\right)^{m}\right), & t_{(3\times1(6))} \leq t \end{cases}$$

(3.7).

Taking into account (3.3), we have
$$\theta_{DO} = 1 - \theta_{\alpha(2\times4)} - \theta_{(3\times1(6))}.$$  (3.8)

It follows from expression (3.4):
$$\frac{\partial I^{SB}}{\partial t} = \Delta I_1 \cdot \frac{\partial \theta_{\alpha(2\times4)}}{\partial t} + \Delta I_2 \cdot \frac{\partial \theta_{(3\times1(6))}}{\partial t},$$  (3.9)

where $\Delta I_1 \equiv I_{\alpha(2\times4)} - I_{DO}$ and $\Delta I_2 \equiv I_{(3\times1(6))} - I_{DO}$.

The fitting of both *SBI* time dependences proper and their derivatives is carried out to determine the structural transition parameters. The main model parameters $t_{(1\times3(6))}$, $\gamma_{DO}$ and $\gamma_{(1\times3(6))}$ were supposed to satisfy the Arrenius law.

The results of the analysis are presented in Fig. 10. It is seen that the model curve correctly describes both the experimental data (Fig. 10 *a*)) and their derivatives on time (Fig. 10 *b*)). The obtained dependences $\theta_{\alpha(2\times4)}$ and $\theta_{(3\times1(6))}$ (Fig. 10 *c*)) allowed us to establish the degrees of superstructural transitions $m^{I} = 2$ and $m^{II} = 2$ (Fig. 11 *a*) and *b*)).

This confirms the correctness of our base statement about the domain structure of the transition.

In Fig. 11 variable $\zeta$ is a natural logarithm from time and $\chi$ is a double natural logarithm from *SBI*. Indices I and II denote superstructural transitions $\alpha(2\times4) \rightarrow DO$ and $DO \rightarrow (3\times1(6))$, respectively. As these degrees coincide with the system's dimensionality ($m = d = 2$ for 2*D*), the transitions are characterized by a fixed number of nuclei of a new phase during the whole process. The analysis of the temperature dependence of the velocity for superstructural transitions $\alpha(2\times4) \rightarrow DO$ and $DO \rightarrow (3\times1(6))$ allowed establishing their activation energy values equal to 3.44 ± 0.08 eV and 3.73 ±0.09 eV, respectively (Fig. 12).

It is also should be noted that the *SBI* dependence on time (Fig. 10 *a*)) has a clearly expressed peculiarity, i.e. minimum. The position of this minimum characterizes the completion of transition $\alpha(2\times4) \to DO$ and it depends only on temperature. This circumstance can be used for the precise determination of GaAs(001) substrates temperature.

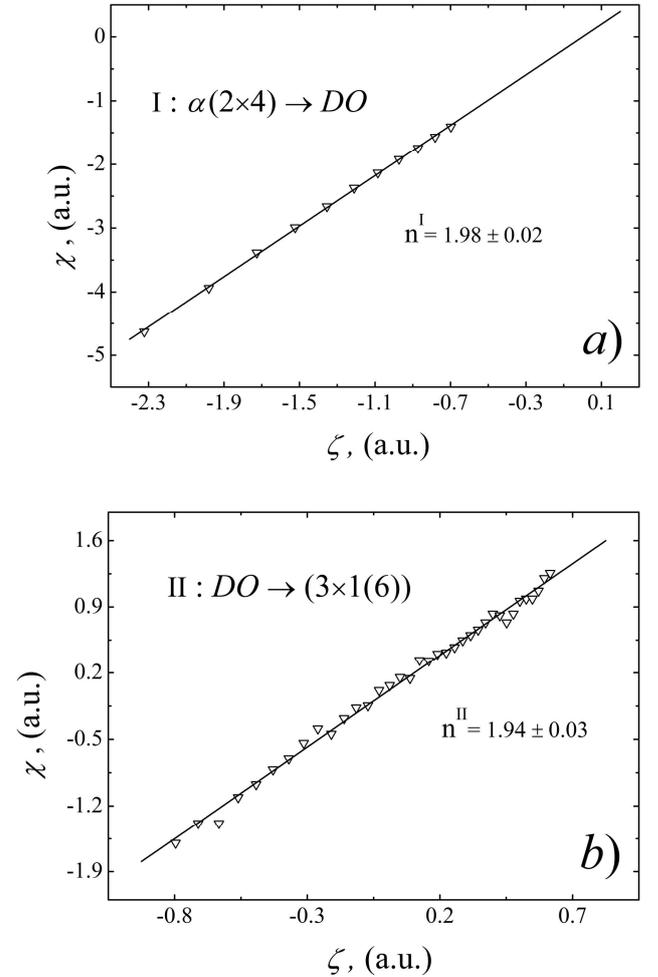

**Fig. 11.** Parameters *n* of *a*) $\alpha(2\times4) \rightarrow DO$ and *b*) $DO \rightarrow (3\times1(6))$ superstructural transitions.
▽ – exp. data; ▬ – linear approximation.

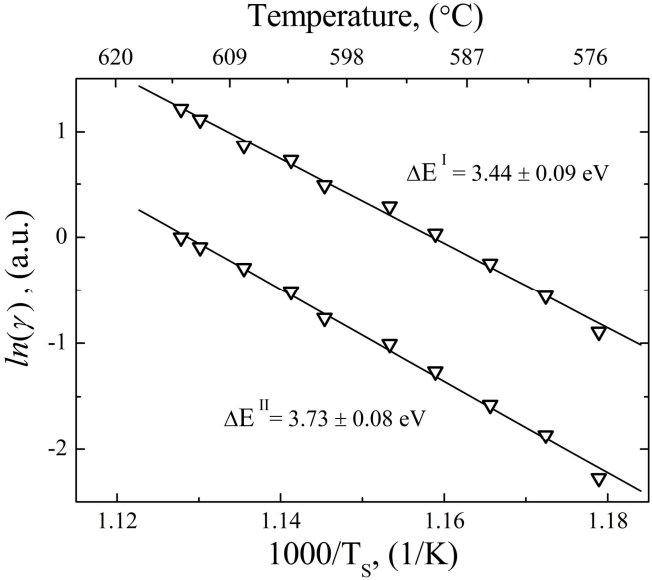

**Fig. 12.** Activation energy values of superstructural transitions $\alpha(2\times4) \to DO$ and $DO \to (3\times1(6))$.
▽ – exp. data; ▬ – linear approximation.

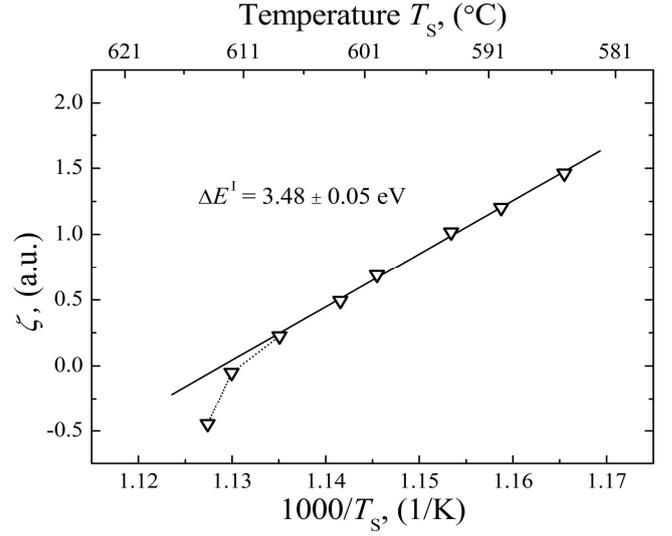

**Fig. 13.** Calibration dependence for the procedure of precise determining the GaAs(001) substrates temperature.
▽ – exp. data; ▬ – linear approximation.

According to the *JMAK* kinetic model, the average time $T$ of the completion of structural $\alpha(2\times4) \to DO$ transition process can be evaluated as a "lifetime" of the starting $\alpha(2\times4)$ phase random point

$$T = \int_{t_0}^{\infty} \theta_{\alpha(2\times4)}(t)dt. \qquad (3.10)$$

Taking into account that
$$\theta_{\alpha(2\times4)}(t) = \exp\left(-\gamma_{\alpha(2\times4)}^d (t - t_{\alpha(2\times4)})^m\right),$$
we will get
$$T = \frac{1}{m \cdot \gamma_{\alpha(2\times4)}^{d/m}} \cdot \Gamma\left(\frac{1}{m}\right), \qquad (3.11)$$

here $\Gamma(z) = \int_0^{\infty} x^{z-1} e^{-x} dx$ – Euler function.

Thus, for the process with $m = d = 2$, we will have
$$T = \frac{\sqrt{\pi}}{2\gamma_{\alpha(2\times4)}}. \qquad (3.12)$$

Hence, $T$ depends on time as $1/\gamma_{\alpha(2\times4)}$.

The moments of the onset (corresponds to the moment of closing the valve of arsenic source) and ending (corresponds to $\frac{\partial I^{SB}}{\partial t} = 0$ – the moment of $I^{SB}$ minimum) of the structural $\alpha(2\times4) \to DO$ transition process are clearly seen on the curves that characterize the derivative of *SBI* dependence on time (Fig. 10 *b*)). It enables us to determine value *T*. According to expression (3.12), the dependence of the natural logarithm of value *T* (variable $\zeta$) on the reverse substrate temperature is a straight line with the inclination angle corresponding to activation energy $\gamma_{\alpha(2\times4)}$. These peculiarities can be used for the procedure of precise determining the GaAs(001) substrates temperature on the time of transition $\alpha(2\times4) \to DO$. This dependence is presented in Fig. 13. One should note that, for temperatures higher than 608°C, the measured value *T* has lower (related to the approximation line) values. It's because, at temperatures higher than 608°C part of the surface is already in the *DO* state, even under the arsenic flux.

## 4. Conclusion

The evolution of GaAs(001) surface structural properties, during the transition $\alpha(2\times4) \to (3\times1(6))$ initiated by a sharp change of the As$_2$ flux, was studied.

It is shown that this transition is a complex of five, changing each other, superstructural states which we interpreted as $\alpha(2\times4)$, *DO*, $(6\times6)$, $(n\times6)$ and $(3\times1)$. The reconstruction $\alpha(2\times4)$ with a big number of As-dimer vacancies is understood by state *DO*. This surface still preserves the short-range order, but it has already lost its long-range order. States $(6\times6)$ and $(n\times6)$ are related (characterized by inconsiderable structural differences). The RHEED rocking curves of the corresponding states differ only in a narrow range of $\alpha$ angles ($1.84 \div 2.07$ angular degrees).

The superstructural transition kinetics in the temperature range $570 \div 614°C$ was analyzed within the *JMAK* (*Johnson – Melh – Avrami – Kolmogorov*) model. It is shown that transitions $\alpha(2\times4) \to DO$ and $DO \to (3\times1(6))$ are realized through the processes of islands (superstructural domains) nucleation, growth and coalescence. These transitions are characterized by a fixed number of domains during the whole process. The activation energies of transitions $\Delta E^{\mathrm{I}} = 3.44 \pm 0.08$ eV and $\Delta E^{\mathrm{II}} = 3.73 \pm 0.09$ eV, respectively, were determined. The procedure for precise determination of GaAs(001) surface temperature using the features of the $\alpha(2\times4) \to DO$ transition process kinetic was proposed.

**Acknowledgement**

This work was supported by the Russian Science Foundation, grant № 16-12-00023.